\documentclass[final]{IEEEtran}

\usepackage[encapsulated]{CJK}
\usepackage{ucs}
\usepackage[utf8x]{inputenc}
\usepackage[cmex10]{amsmath}
\usepackage{amsmath,amssymb,amscd,bbm,amsthm,mathrsfs,dsfont}
\usepackage{algorithmic,algorithm}
\usepackage{mdwmath}
\usepackage{mdwtab}
\usepackage{bm,upgreek}
\usepackage{cite}
\usepackage{rotating,graphics,psfrag,epsfig}
\usepackage{array}
\usepackage{booktabs}
\usepackage{indentfirst}
\usepackage{subfigure}
\usepackage{lipsum,fancyhdr,lastpage,refcount}
\usepackage[hyphens]{url}
\usepackage{fixltx2e,dblfloatfix}
\usepackage{blindtext}
\usepackage[T1]{fontenc}
\usepackage{color}

\IEEEoverridecommandlockouts

\let\oldnl\nl
\newcommand{\nonl}{\renewcommand{\nl}{\let\nl\oldnl}}

\begin{document}

\title{Secrecy Preserving in Stochastic Resource Orchestration for Multi-Tenancy Network Slicing}

%\author{\IEEEauthorblockN{Xianfu Chen, Zhifeng Zhao, Celimuge Wu, Tao Chen, and Honggang Zhang}

%\thanks{X. Chen and T. Chen are with the VTT Technical Research Centre of Finland, Oulu, Finland (e-mail: \{xianfu.chen, tao.chen\}@vtt.fi). Z. Zhao and H. Zhang are with the College of Information Science and Electronic Engineering, Zhejiang University, Hangzhou, China (e-mail: \{zhaozf, honggangzhang\}@zju.edu.cn). C. Wu is with the Graduate School of Informatics and Engineering, University of Electro-Communications, Tokyo, Japan (email: clmg@is.uec.ac.jp).}
%}

\author{\IEEEauthorblockN{Xianfu Chen\IEEEauthorrefmark{1}, Zhifeng Zhao\IEEEauthorrefmark{2}, Celimuge Wu\IEEEauthorrefmark{3}, Tao Chen\IEEEauthorrefmark{1}, Honggang Zhang\IEEEauthorrefmark{2}, and Mehdi Bennis\IEEEauthorrefmark{4}}

{\IEEEauthorrefmark{1}VTT Technical Research Centre of Finland Ltd, Finland}\\
{\IEEEauthorrefmark{2}College of Information Science and Electronic Engineering, Zhejiang University, China}\\
{\IEEEauthorrefmark{3}Graduate School of Informatics and Engineering, University of Electro-Communications, Tokyo, Japan}\\
{\IEEEauthorrefmark{4}Centre for Wireless Communications, University of Oulu, Finland}
\vspace{-.8cm}
}

\maketitle

\begin{abstract}

Network slicing is a proposing technology to support diverse services from mobile users (MUs) over a common physical network infrastructure.
In this paper, we consider radio access network (RAN)-only slicing, where the physical RAN is tailored to accommodate both computation and communication functionalities.
Multiple service providers (SPs, i.e., multiple tenants) compete with each other to bid for a limited number of channels across the scheduling slots, aiming to provide their subscribed MUs the opportunities to access the RAN slices.
An eavesdropper overhears data transmissions from the MUs.
We model the interactions among the non-cooperative SPs as a stochastic game, in which the objective of a SP is to optimize its own expected long-term payoff performance.
To approximate the Nash equilibrium solutions, we first construct an abstract stochastic game using the channel auction outcomes.
Then we linearly decompose the per-SP Markov decision process to simplify the decision-makings and derive a deep reinforcement learning based scheme to approach the optimal abstract control policies.
TensorFlow-based experiments verify that the proposed scheme outperforms the three baselines and yields the best performance in average utility per MU per scheduling slot.

\end{abstract}

\section{Introduction}
\label{intr}

To keep up with the proliferation of wireless services, new cell sites are being constantly built, eventually leading to dense network deployments \cite{Andr12}.
However, it becomes extremely complex to operate the control plane functions in a dense radio access network (RAN).
In recent years, the computation-intensive applications (e.g., augmented reality and interactive online gaming) are gaining increasing popularity \cite{Mao17}.
The mobile user (MU)-end terminal devices are in general constrained by battery capacity and processing speed of the central processing unit (CPU).
The tension between computation-intensive applications and resource-constrained terminal devices calls for a revolution in computing \cite{Saty17}.
Mobile-edge computing (MEC) is envisioned as a promising solution, which brings the computing capabilities within the RANs in close proximity to MUs \cite{Mao17}.
Offloading a computation task to a MEC server for execution involves data transmissions.
How to orchestrate radio resources between MEC and traditional mobile services adds another dimension of complexity to the network operations \cite{Zhou17}.
By abstracting all physical base stations (BSs) in a geographical area as a logical big BS, the software-defined networking (SDN) concept provides infrastructure flexibility as well as service-oriented customization \cite{Gudi13}.
In a software-defined RAN, the SDN-orchestrator handles all control plane operations.

One key benefit from a software-defined RAN is to facilitate network sharing \cite{Lian15}.
As such, the same physical network is able to host multiple service providers (SPs, namely, multiple tenants \cite{Xiao1802}), which breaks the traditional business model regarding the single ownership of a network infrastructure \cite{Fris08}.
For example, an over-the-top application provider (e.g., Google \cite{goog18}) can be a SP so as to lease radio resources from the infrastructure provider to improve the Quality-of-Service and the Quality-of-Experience for its subscribers.
Building upon the 3GPP TSG SA 5 network sharing paradigm \cite{3gpp18}, a software-defined RAN architecture and its integration with network function virtualization enable RAN-only slicing that splits the RAN into multiple virtual slices \cite{Sall17}.
This paper is primarily concerned with a software-defined RAN where the RAN slices are specifically tailored to accommodate both computation and communication functionalities \cite{Shah17}.

The technical challenges yet remain for the implementation of RAN-only slicing.
Particularly, the mechanisms that exploit the decoupling of control and data planes in a software-defined RAN must be developed to optimize radio resource utilization.
For the considered software-defined RAN, a limited number of channels are auctioned over the time horizon to the MUs, which request MEC and traditional mobile services.
An eavesdropper exists in the network and overhears the MUs during the data transmissions \cite{Wu16}.
Multiple SPs compete to orchestrate the channels for their subscribed MUs according to the network dynamics, aiming to maximize the expected long-term payoff performance.
Upon receiving the auction bids from all SPs, the SDN-orchestrator allocates channels to the MUs through a Vickrey-Clarke-Groves (VCG) mechanism\footnote{The VCG mechanism ensures truthfulness, efficiency and incentive compatibility.} \cite{Ji07}.
To combat the threat from the eavesdropper, each MU then proceeds to use a secrecy-rate \cite{Wu16} to offload computation tasks and schedule packets over the assigned channel.
To the best of our knowledge, there does not exist a comprehensive study on stochastic resource orchestration in multi-tenancy RAN-only slicing with secrecy preserving.

\section{System Model}
\label{sysm}

\begin{figure}[t]
  \centering
  \includegraphics[width=20pc]{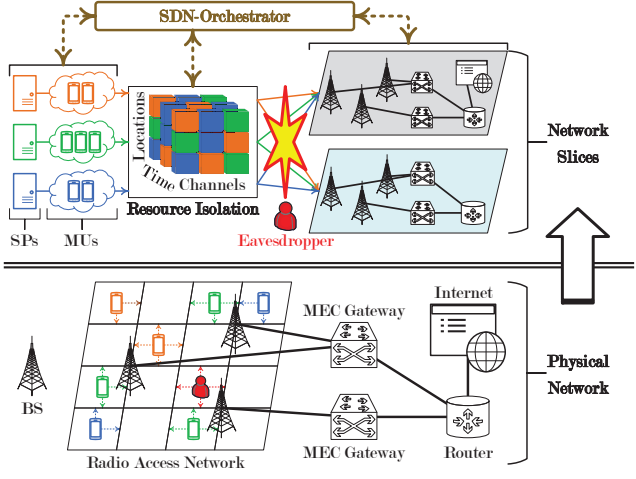}
  \caption{Illustration of the RAN-only slicing architecture.
  An eavesdropper overhears the data transmissions from the MUs across the time horizon.}
  \label{systMode}
\end{figure}
As shown in Fig. \ref{systMode}, we focus on a system model with RAN-only slicing, where an eavesdropper intentionally overhears the data transmissions of the MUs.
The time horizon is divided into discrete scheduling slots, each of which is indexed by an integer $k \in \mathds{N}_+$ and is assumed to be of equal duration $\delta$ (in seconds).
The RAN consists of a set $\mathcal{B}$ of physical BSs covering a service area, which can be represented by a set $\mathcal{L}$ of small locations with each being characterized by uniform signal propagation conditions \cite{Chen15}.
We use $\mathcal{L}_b$ to denote the serving area of a BS $b \in B$.
For any two BSs $b$ and $b' \in \mathcal{B}$ ($b' \neq b$), we assume that $\mathcal{L}_b \cap \mathcal{L}_{b'} = \emptyset$.
We denote the geographical distribution of BSs by a topological graph $\mathcal{TG} = \langle\mathcal{B}, \mathcal{E}\rangle$, where $\mathcal{E} = \{e_{b, b'}: b \neq b', b, b' \in \mathcal{B}\}$ with $e_{b, b'} = 1$ if BSs $b$ and $b'$ are neighbours and otherwise $e_{b, b'} = 0$.
Suppose that $I$ SPs provide both MEC and traditional mobile services to MUs while each MU can subscribe to only one SP.
Let $\mathcal{N}_i$ be the set of MUs of a SP $i \in \mathcal{I} = \{1, \cdots, I\}$.

Across the scheduling slots, the MUs and the eavesdropper move within $\mathcal{L}$ following a Markov mobility model \cite{Nich08}.
Denote by $\mathcal{N}_{b, i}^k$ the set of MUs of SP $i \in \mathcal{I}$ moving into the area of a BS $b \in \mathcal{B}$ during a slot $k$.
We assume that a MU at a location can only be associated with the BS that covers the location.
In the network, all MUs share a set $\mathcal{J} = \{1, \cdots, J\}$ of orthogonal channels with the same bandwidth $\eta$ (in Hz).
The SPs compete for the limited channel access opportunities for their MUs.
Specifically, at the beginning of a scheduling slot $k$, each SP $i$ submits an auction bid $\bm\beta_i^k = (\nu_i^k, \mathbf{C}_i^k)$, where $\nu_i^k$ is the valuation over $\mathbf{C}_i^k = (C_{b, i}^k: b \in \mathcal{B})$ with $C_{b, i}^k$ being the number of requested channels in the service area of a BS $b$.
After receiving $\bm\beta^k = (\bm\beta_i^k: i \in \mathcal{I})$, the SDN-orchestrator performs channel allocation and calculates payment $\tau_i^k$ for each SP $i$.
Let $\bm\rho_n^k = (\rho_{n, j}^k: j \in \mathcal{J})$ be the channel allocation of a MU $n \in \mathcal{N} = \cup_{i \in \mathcal{I}} \mathcal{N}_i$, where $\rho_{n, j}^k = 1$ if channel $j$ is allocated to MU $n\in \mathcal{N}$ during slot $k$ and $\rho_{n, j}^k = 0$, otherwise.
We also apply the following constraints for centralized channel allocation at the SDN-orchestrator during a slot,
\begin{align}
        \Bigg(\sum_{i \in \mathcal{I}} \sum_{n \in \mathcal{N}_{b, i}^k}  \rho_{n, j}^k\Bigg) \cdot
 & \!   \Bigg(\sum_{i \in \mathcal{I}} \sum_{n \in \mathcal{N}_{b', i}^k} \rho_{n, j}^k\Bigg) = 0,          \nonumber\\
        \mbox{if } e_{b, b'}
 & =    1, \forall e_{b, b'} \in \mathcal{E}, \forall j \in \mathcal{J};                                    \label{c1}\\
        \sum_{i \in \mathcal{I}} \sum_{n \in \mathcal{N}_{b, i}^k} \rho_{n, j}^k
 & \leq 1, \forall b \in \mathcal{B}, \forall j \in \mathcal{J};                                            \label{c2}\\
        \sum_{j \in \mathcal{J}} \rho_{n, j}^k
 & \leq 1, \forall b \in \mathcal{B}, \forall i \in \mathcal{I}, \forall n \in \mathcal{N}_{b, i},          \label{c3}
\end{align}
which ensure that one channel cannot be allocated to MUs associated with two adjacent BSs in order to avoid interference during data transmissions, while in the service area of a BS, one MU can be assigned at most one channel and one channel can be assigned to at most one MU.
Denote $\bm\phi^k = (\phi_i^k: i \in \mathcal{I})$ as the winner vector at the beginning of a scheduling slot $k$, where $\phi_i^k = 1$ if SP $i$ wins the channel auction and $\phi_i^k = 0$ indicates that no channel will be allocated to the MUs of SP $i$ during the slot.
The SDN-orchestrator determines $\bm\phi^k$ via the VCG pricing mechanism, namely,
\begin{equation}\label{chanSche}
  \begin{array}{cl}
                  & \bm\phi^k = \underset{\bm\phi}{\arg\max} \displaystyle\sum\limits_{i \in \mathcal{I}} \phi_i \cdot \nu_i^k    \\
    \mathrm{s.t.} & \mbox{constraints (\ref{c1}), (\ref{c2}) and (\ref{c3})};                                                           \\
                  & \displaystyle\sum_{n \in \mathcal{N}_{b, i}^k} \varphi_n^k = \phi_i \cdot C_{b, i}^k,
                    \forall b \in \mathcal{B}, \forall i \in \mathcal{I},
  \end{array}
\end{equation}
where $\varphi_n^k = \sum_{j \in \mathcal{J}} \rho_{n, j}^k$ and $\bm\phi = (\phi_i: i \in \mathcal{I})$ with $\phi_i \in \{0, 1\}$.
The payment $\tau_i^k$ for each SP $i$ can be calculated to be $\tau_i^k = \max_{\bm\phi_{-i}} \sum_{i' \in \mathcal{I} \setminus \{i\}} \phi_{i'} \cdot \nu_{i'}^k - \max_{\bm\phi} \sum_{i' \in \mathcal{I} \setminus \{i\}} \phi_{i'} \cdot \nu_{i'}^k$, where $-i$ denotes all the competitors of SP $i$.

Let $L_{n, (\mathrm{u})}^k$ and $L_{(\mathrm{e})}^k \in \mathcal{L}$ be the geographical locations of a MU $n \in \mathcal{N}$ and the eavesdropper during a scheduling slot $k$, respectively.
As in \cite{Chen15}, we assume that the average channel gains $H_{n, (\mathrm{u})}^k = h_{(\mathrm{u})}(L_{n, (\mathrm{u})}^k)$ and $H_{n, (\mathrm{e})}^k = h_{(\mathrm{e})}(L_{n, (\mathrm{u})}^k, L_{(\mathrm{e})}^k)$ of links between MU $n$ and the associated BS as well as the eavesdropper are determined by the respective distances.
At the beginning of each scheduling slot $k$, MU $n$ independently generates a random number $A_{n, (\mathrm{t})}^k \in \mathcal{A} = \{0, 1, \cdots, A_{(\mathrm{t})}^{(\max)}\}$ of computation tasks\footnote{To ease analysis, we assume that the maximum CPU power at a mobile device matches the maximum computation task arrivals and a MU can process $A_{(\mathrm{t})}^{(\max)}$ tasks within one scheduling slot.} according to a Markov process \cite{He17}.
We represent a computation task by $(\mu_{(\mathrm{t})}, \vartheta)$, where $\mu_{(\mathrm{t})}$ and $\vartheta$ are, respectively, the input data size (in bits) and the number of CPU cycles required to accomplish one input bit of the computation task.
For a computation task, two decisions are available: 1) to be processed locally at the MU; or 2) to be offloaded to the MEC server in the computation slice for execution.
The computation offloading decision for MU $n$ at a slot $k$ specifies the number $R_{n, (\mathrm{t})}^k$ of tasks to be transmitted to the MEC server.
Then the remaining $A_{n, (\mathrm{t})}^k - \varphi_n^k \cdot R_{n, (\mathrm{t})}^k$ tasks are to be processed locally.
Meanwhile, a data queue at a MU buffers the packets from the traditional mobile service.
Let $W_n^k$ and $A_{n, (\mathrm{p})}^k$ be the queue length and the random new packet arrivals for MU $n$ at the beginning of a slot $k$.
We assume that the data packets are of a constant size $\mu_{(\mathrm{p})}$ (bits) and the packet arrival process is independent among the MUs and identical and independently distributed across time.
Let $R_{n, (\mathrm{p})}^k$ be the number of packets that are scheduled for transmission from MU $n$ at scheduling slot $k$.
The queue evolution of MU $n$ can be written as the form below,
\begin{align}
  W_n^{k + 1} = \min\!\left\{W_n^k - \varphi_n^k \cdot R_{n, (\mathrm{p})}^k + A_{n, (\mathrm{p})}^k, W^{(\max)}\right\},
\end{align}
where $W^{(\max)}$ is the queue length limit.

%-----------------------------------------------------------
To ensure security, the energy (in Joules) consumed by a MU $n \in \mathcal{N}$ for transmitting $\varphi_n^k \cdot R_{n, (\mathrm{t})}^k$ computation tasks and $\varphi_n^k \cdot R_{n, (\mathrm{p})}^k$ data packets with a secrecy-rate \cite{Wu16} during a slot $k$ can be calculated as
\begin{align}\label{tranPowe}
 & P_{n, (\mathrm{tr})}^k =                                                                                                                             \\
 & \left\{\!\!
   \begin{array}{l@{~}l}
     \frac{\delta \cdot \eta \cdot \sigma^2 \cdot
        \bigg(2^{\frac{\varphi_n^k \cdot \left(\mu_{(\mathrm{t})} \cdot R_{n, (\mathrm{t})}^k + \mu_{(\mathrm{p})} \cdot R_{n, (\mathrm{p})}^k\right)}
        {\eta \cdot \delta}} - 1\bigg)}
        {H_{n, (\mathrm{u})}^k - H_{(\mathrm{e})}^k \cdot
        2^{\frac{\varphi_n^k \cdot \left(\mu_{(\mathrm{t})} \cdot R_{n, (\mathrm{t})}^k + \mu_{(\mathrm{p})} \cdot R_{n, (\mathrm{p})}^k\right)}
        {\eta \cdot \delta}}},                                                    & \mbox{if } H_{n, (\mathrm{u})}^k > H_{(\mathrm{e})}^k;              \\
   0,                                                                             & \mbox{otherwise},
   \end{array}
   \right.                                                                                                                                              \nonumber
\end{align}
where $\sigma^2$ is the background noise power spectral density.
Let $\Omega^{(\max)}$ be the maximum transmit power for all MUs, namely, $P_{n, (\mathrm{tr})}^k \leq \Omega^{(\max)} \cdot \delta$, $\forall n$ and $\forall k$.
For the remaining $A_{n, (\mathrm{t})}^k - \varphi_n^k \cdot R_{n, (\mathrm{t})}^k$ computation tasks that are to be locally processed, the CPU energy consumption is
\begin{align}\label{taskEner}
  P_{n, (\mathrm{CPU})}^k = \varsigma \cdot \mu_{(\mathrm{t})} \cdot \vartheta \cdot \varrho^2 \cdot \left(A_{n, (\mathrm{t})}^k - \varphi_n^k \cdot R_{n, (\mathrm{t})}^k\right),
\end{align}
where $\varsigma$ is the effective switched capacitance \cite{Burd96} and $\varrho$ is the CPU-cycle frequency of the MU-end devices.

\section{Stochastic Game Formulation}

%-----------------------------------------------------------
At a scheduling slot $k$, the local state of a MU $n \in \mathcal{N}$ is described as $\bm\chi_n^k = (L_{n, (\mathrm{u})}^k, L_{(\mathrm{e})}^k, A_{n, (\mathrm{t})}^k, W_n^k) \in \mathcal{X} = \mathcal{L}^2 \times$ $\mathcal{A} \times \mathcal{W}$, where the SDN-orchestrator broadcasts the information of $L_{(\mathrm{e})}^k$ to all MUs.
Then $\bm\chi^k = (\bm\chi_n^k: n \in \mathcal{N}) \in \mathcal{X}^{|\mathcal{N}|}$ characterizes the global network state, where $|\mathcal{N}|$ means the cardinality of the set $\mathcal{N}$.
Define by $\bm\pi_i = (\pi_{i, (\mathrm{c})}, \bm\pi_{i, (\mathrm{t})}, \bm\pi_{i, (\mathrm{p})})$ a control policy of a SP $i \in \mathcal{I}$, where $\pi_{i, (\mathrm{c})}$, $\bm\pi_{i, (\mathrm{t})} = (\pi_{n, (\mathrm{t})}: n \in \mathcal{N}_i)$ and $\bm\pi_{i, (\mathrm{p})} = (\pi_{n, (\mathrm{p})}: n \in \mathcal{N}_i)$ are the channel auction, the computation offloading and the packet scheduling policies, respectively.
%
%Note that the computation offloading policy $\pi_{n, (\mathrm{t})}$ as well as the packet scheduling policy $\pi_{n, (\mathrm{p})}$ are MU-specified, hence both $\bm\pi_{i, (\mathrm{t})}$ and $\bm\pi_{i, (\mathrm{p})}$ depend only on $\bm\chi_i^k = (\bm\chi_n^k: n \in \mathcal{N}_i) \in \mathcal{X}_i = \mathcal{X}^{|\mathcal{N}_i|}$.
%
The joint control policy of all SPs is given by $\bm\pi = (\bm\pi_i: i \in \mathcal{I})$.
With the observation of $\bm\chi^k$ at the beginning of each scheduling slot $k$, SP $i$ announces the auction bid $\bm\beta_i^k$ to the SDN-orchestrator and decides the $\mathbf{R}_{i, (\mathrm{t})}^k$ computation tasks as well as $\mathbf{R}_{i, (\mathrm{p})}^k$ packets to be transmitted following $\bm\pi_i$.
That is, $\bm\pi_i(\bm\chi^k) = (\pi_{i, (\mathrm{c})}(\bm\chi^k), \bm\pi_{i, (\mathrm{t})}(\bm\chi_i^k), \bm\pi_{i, (\mathrm{p})}(\bm\chi_i^k)) = (\bm\beta_i^k, \mathbf{R}_{i, (\mathrm{t})}^k, \mathbf{R}_{i, (\mathrm{p})}^k)$, where $\mathbf{R}_{i, (\mathrm{t})}^k = (R_{n, (\mathrm{t})}^k: n \in \mathcal{N}_i)$ and $\mathbf{R}_{i, (\mathrm{p})}^k = (R_{n, (\mathrm{p})}^k: n \in \mathcal{N}_i)$.
Accordingly, SP $i$ realizes an instantaneous payoff
\begin{align}\label{payOff}
 & F_i\!\left(\bm\chi^k, \bm\varphi_i^k, \mathbf{R}_{i, (\mathrm{t})}^k, \mathbf{R}_{i, (\mathrm{p})}^k\right)                                      \nonumber\\
 & = \sum_{n \in \mathcal{N}_i} \alpha_n \cdot U_n\!\left(\bm\chi_n^k, \varphi_n^k, R_{n, (\mathrm{t})}^k, R_{n, (\mathrm{p})}^k\right) - \tau_i^k,
\end{align}
where $\bm\varphi_i^k = (\varphi_n^k: n \in \mathcal{N}_i)$ and $\alpha_n \in \mathds{R}_+$ is the unit price to charge a MU $n$ for achieving utility
\begin{align}\label{util}
  &   U_n\!\left(\bm\chi_n^k, \varphi_n^k, R_{n, (\mathrm{t})}^k, R_{n, (\mathrm{p})}^k\right)
    = U_n^{(1)}\!\left(W_n^{k + 1}\right) + U_n^{(2)}\!\left(D_n^k\right)                                                           \nonumber\\
  & + \ell_n \cdot \left(U_n^{(3)}\!\left(P_{n, (\mathrm{CPU})}^k\right) + U_n^{(4)}\!\left(P_{n, (\mathrm{tr})}^k\right)\right).
\end{align}
In (\ref{util}), $D_n^k = \max\{W_n^k - \varphi_n^k \cdot R_{n, (\mathrm{p})}^k + A_{n, (\mathrm{p})}^k - W^{(\max)}, 0\}$ defines the number of packet drops,
$U_n^{(1)}(\cdot)$, $U_n^{(2)}(\cdot)$, $U_n^{(3)}(\cdot)$ and $U_n^{(4)}(\cdot)$ are the positive and monotonically decreasing functions,
and $\ell_n \in \mathds{R}_+$ is a weighting factor.
Obviously, the randomness lying in $\{\bm\chi^k: k \in \mathds{N}_+\}$ is Markovian.

%-----------------------------------------------------------
Taking expectation with respect to the sequence of per-slot instantaneous payoffs, the expected long-term payoff of a SP $i \in \mathcal{I}$ for a given initial global network state $\bm\chi^1 = \bm\chi \triangleq (\bm\chi_n = (L_{n, (\mathrm{u})}, L_{(\mathrm{e})}, A_{n, (\mathrm{t})}, W_n): n \in \mathcal{N})$ can be expressed as in (\ref{expePayo}),
\begin{figure*}[!t]
\vspace{-.8cm}
\begin{align}\label{expePayo}
  V_i\!\left(\bm\chi, \bm\pi\right) = (1 - \gamma) \cdot
  \textsf{E}_{\bm\pi}\!\!\!\left[\sum_{k = 1}^{\infty} (\gamma)^{k - 1} \cdot
  F_i\!\left(\bm\chi^k, \bm\varphi_i\!\left(\bm\pi_{(\mathrm{c})}\!\left(\bm\chi^k\right)\right), \bm\pi_{i, (\mathrm{t})}\!\left(\bm\chi_i^k\right),
  \bm\pi_{i, (\mathrm{p})}\!\left(\bm\chi_i^k\right)\right) | \bm\chi^1 = \bm\chi\right]
\end{align}
\vspace{-.2cm}
\hrule
\end{figure*}
where $\gamma \in [0, 1)$ is a discount factor.
$V_i(\bm\chi, \bm\pi)$ is also termed as the state-value function of SP $i$.
The aim of each SP $i$ is to device a best-response control policy $\bm\pi_i^*$ such that $\bm\pi_i^* = \arg\max_{\bm\pi_i} V_i(\bm\chi, \bm\pi_i, \bm\pi_{-i})$, $\forall \bm\chi \in \mathcal{X}^{|\mathcal{N}|}$.
Due to the limited number of channels and the stochastic nature in networking environment, we formulate the interactions among multiple non-cooperative SPs over the scheduling slots as a stochastic game, $\mathcal{SG}$, in which $I$ SPs are the players and there are a set $\mathcal{X}^{|\mathcal{N}|}$ of global network states and a collection of control policies $\{\bm\pi_i: \forall i \in \mathcal{I}\}$.
A Nash equilibrium (NE), which is a tuple of control policies $\langle \bm\pi_i^*: i \in \mathcal{I}\rangle$, describes the rational behaviours of the SPs in a $\mathcal{SG}$.
For the $I$-player $\mathcal{SG}$ with expected infinite-horizon discounted payoffs, there always exists a NE in stationary control policies \cite{Fink64}.
Define $\mathds{V}_i(\bm\chi) = V_i(\bm\chi, \bm\pi_i^*, \bm\pi_{-i}^*)$ as the optimal state-value function, $\forall i \in \mathcal{I}$ and $\forall \bm\chi \in \mathcal{X}^{|\mathcal{N}|}$.

\section{Abstract Stochastic Game Reformulation and Deep Reinforcement Learning}
\label{probSolv}

From (\ref{expePayo}), it can be easily observed that the expected long-term payoff of a SP $i \in \mathcal{I}$ depends on information of not only the global network state across the scheduling slots but also the joint control policy $\bm\pi$.
In other words, the decision makings from the non-cooperative SPs are coupled in the $\mathcal{SG}$, which makes it a challenging task to find the NE.
In this section, we elaborate on how the SPs play the $\mathcal{SG}$ only with limited local information.

\subsection{Stochastic Game Abstraction}
\label{SGA}

%-----------------------------------------------------------
To capture the coupling of decision makings among the SPs, we abstract $\mathcal{SG}$ as $\mathcal{AG}$ \cite{Chen1801}, in which a SP $i \in \mathcal{I}$ behaves based on its own local network dynamics and abstractions of states at other competing SPs.
Let $\mathcal{S}_i = \{1, \cdots, S_i\}$ be an abstraction of the state space $\mathcal{X}_{-i}$, where $S_i \in \mathds{N}_+$ and $S_i \ll |\mathcal{X}_{-i}|$.
We observe that the behavioural couplings in $\mathcal{SG}$ exist in the channel auction and the payments of SP $i$ depend on $\mathcal{X}_{-i}$.
This allows SP $i$ to construct $\mathcal{S}_i$ by classifying the value region $[0, \Gamma_i]$ of payments into $S_i$ intervals, i.e., $[0, \Gamma_{i, 1}]$, $(\Gamma_{i, 1}, \Gamma_{i, 2}]$, $(\Gamma_{i, 2}, \Gamma_{i, 3}]$, $\ldots$, $(\Gamma_{i, S_i - 1}, \Gamma_{i, S_i}]$, where $\Gamma_{i, S_i} = \Gamma_i$ is the maximum payment and we let $\Gamma_{i, 1} = 0$ for a special case in which SP $i$ wins the channel auction with no payment\footnote{This case happens when there are enough channels to serve all MUs in the network \cite{Jia09}.}.
With this regard, SP $i$ abstracts $(\bm\chi_i, \bm\chi_{- i}) \in \mathcal{X}^{|\mathcal{N}|}$ as $\tilde{\bm\chi}_i = (\bm\chi_i, s_i) \in \tilde{\mathcal{X}}_i= \mathcal{X}_i \times \mathcal{S}_i$ if the payment in previous scheduling slot belongs to $(\Gamma_{i, s_i - 1}, \Gamma_{i, s_i}]$.

%-----------------------------------------------------------
Let $\tilde{\bm\pi}_i = (\tilde{\pi}_{i, (\mathrm{c})}, \bm\pi_{i, (\mathrm{t})}, \bm\pi_{i, (\mathrm{p})})$ be the abstract control policy in the $\mathcal{AG}$ played by a SP $i \in \mathcal{I}$ over $\tilde{\mathcal{X}}_i$, where $\tilde{\pi}_{i, (\mathrm{c})}$ is the abstract channel auction policy.
Likewise, the abstract state-value function for SP $i$ under $\tilde{\bm\pi} = (\tilde{\bm\pi}_i: i \in \mathcal{I})$ can then be defined as in (\ref{abstVFunc}),
\begin{figure*}[!t]
\vspace{-.6cm}
\begin{align}\label{abstVFunc}
       \tilde{V}_i\!\left(\tilde{\bm\chi}_i, \tilde{\bm\pi}\right)
    = (1 - \gamma) \cdot \textsf{E}_{\tilde{\bm\pi}}\!\!\!\left[\sum_{k = 1}^{\infty} (\gamma)^{k - 1} \cdot
      \tilde{F}_i\!\left(\tilde{\bm\chi}_i^k, \bm\varphi_i\!\left(\tilde{\bm\pi}_{(\mathrm{c})}\!\left(\tilde{\bm\chi}^k\right)\right), \bm\pi_{i, (\mathrm{t})}\!\left(\bm\chi_i^k\right), \bm\pi_{i, (\mathrm{p})}\!\left(\bm\chi_i^k\right)\right) | \tilde{\bm\chi}_i^1 = \tilde{\bm\chi}_i\right]
\end{align}
\vspace{-.2cm}
\hrule
\vspace{-.6cm}
\end{figure*}
$\forall \tilde{\bm\chi}_i \in \tilde{\mathcal{X}}_i$, where $\tilde{\bm\chi}^k = (\tilde{\bm\chi}_i^k = (\bm\chi_i^k, s_i^k): i \in \mathcal{I})$ with $s_i^k$ being the abstract state at slot $k$ and $\tilde{F}_i(\tilde{\bm\chi}_i^k, \bm\varphi_i(\tilde{\bm\pi}_{(\mathrm{c})}(\tilde{\bm\chi}^k)), \bm\pi_{i, (\mathrm{t})}(\bm\chi_i^k), \bm\pi_{i, (\mathrm{p})}(\bm\chi_i^k))$ is the immediate payoff with $\tilde{\bm\chi}^k = (\tilde{\bm\chi}_i^k: i \in \mathcal{I})$ and $\tilde{\bm\pi}_{(\mathrm{c})} = (\tilde{\pi}_{i, (\mathrm{c})}: i \in \mathcal{I})$.
In our previous work \cite{Chen1801}, we have proved that instead of playing the original $\bm\pi^*$ in the $\mathcal{SG}$, the NE joint abstract control policy given by $\tilde{\bm\pi}^* = (\tilde{\bm\pi}_i^*: i \in \mathcal{I})$ in the $\mathcal{AG}$ leads to a bounded regret, where $\tilde{\bm\pi}_i^* = (\tilde{\pi}_{i, (\mathrm{c})}^*, \bm\pi_{i, (\mathrm{t})}^*, \bm\pi_{i, (\mathrm{p})}^*)$ denotes the best-response abstract control policy of SP $i$.
Hereinafter, we switch our focus to the $\mathcal{AG}$, in which a SP solves a single-agent Markov decision process (MDP).
Suppose all SPs play $\tilde{\bm\pi}^*$ in the $\mathcal{AG}$.
Denote $\tilde{\mathds{V}}_i(\tilde{\bm\chi}_i) = \tilde{V}_i(\tilde{\bm\chi}_i, \tilde{\bm\pi}^*)$.

\subsection{Decomposition of Abstract State-Value Function}
\label{VDeco}

There remain two challenges involved in solving the optimal abstract state-value functions for each SP $i \in \mathcal{I}$ using dynamic programming methods \cite{Rich98}: 1) a priori knowledge of the abstract network state transition probability is not feasible; and 2) the size of the decision making space $\{\tilde{\bm\pi}_i(\tilde{\bm\chi}_i): \tilde{\bm\chi}_i \in \tilde{\mathcal{X}}_i\}$ grows exponentially as $|\mathcal{N}_i|$ increases.
On the other hand, the channel auction decisions and the computation offloading as well as packet scheduling decisions are made in sequence and are independent across a SP and its subscribed MUs.
We are hence motivated to decompose the per-SP MDP in the $\mathcal{AG}$ into $|\mathcal{N}_i| + 1$ independent MDPs.
More specifically, for a SP $i \in \mathcal{I}$, $\tilde{\mathds{V}}_i(\tilde{\bm\chi}_i)$, $\forall \tilde{\bm\chi}_i \in \tilde{\mathcal{X}}_i$, can be computed as
\begin{equation}\label{VFactDeco}
   \tilde{\mathds{V}}_i\!\left(\tilde{\bm\chi}_i\right)
 = \sum\limits_{n \in \mathcal{N}_i} \alpha_n \cdot \mathds{U}_n(\bm\chi_n) - \mathds{U}_i(s_i),
\end{equation}
where the per-MU $\mathds{U}_n$ and the $\mathds{U}_i(s_i)$ of SP $i$ satisfy, respectively, (\ref{MTBell})
\begin{figure*}[!t]
\vspace{-.8cm}
\begin{align}\label{MTBell}
 &   \mathds{U}_n(\bm\chi_n) =                                                                                                              \\
 &   \max_{R_{n, (\mathrm{t})}, R_{n, (\mathrm{p})}} \!\!\left\{\!(1 - \gamma) \cdot
     U_n\!\left(\bm\chi_n, \varphi_n\!\left(\tilde{\bm\pi}_{(\mathrm{c})}^*\!\left(\tilde{\bm\chi}\right)\right),
     R_{n, (\mathrm{t})}, R_{n, (\mathrm{p})}\right)
   + \gamma \cdot \sum_{\bm\chi_n' \in \mathcal{X}}
     \mathbb{P}\!\left(\bm\chi_n' | \bm\chi_n, \varphi_n\!\left(\tilde{\bm\pi}_{(\mathrm{c})}^*\!\left(\tilde{\bm\chi}\right)\right),
     R_{n, (\mathrm{t})}, R_{n, (\mathrm{p})}\right) \cdot \mathds{U}_n\!\left(\bm\chi_n'\right)\!\right\}                                  \nonumber
\end{align}
\vspace{-.2cm}
\hrule
\end{figure*}
and
\begin{align}\label{WSPBell}
 & \mathds{U}_i(s_i) =                                                                                                              \\
 & (1 - \gamma) \cdot \tau_i +
   \gamma \cdot \sum_{s_i' \in \mathcal{S}_i} \mathbb{P}\!\left(s_i' | s_i,
   \phi_i\!\left(\tilde{\bm\pi}_{(\mathrm{c})}^*\!\left(\tilde{\bm\chi}\right)\right)\right) \cdot \mathds{U}_i\!\left(s_i'\right). \nonumber
\end{align}
In the above, $\tilde{\bm\pi}_{(\mathrm{c})}^*(\tilde{\bm\chi}) = (\tilde{\pi}_{i, (\mathrm{c})}^*(\tilde{\bm\chi}_i): i \in \mathcal{I})$, while $R_{n, (\mathrm{t})}$ and $R_{n, (\mathrm{p})}$ are the computation offloading and packet scheduling decisions under $\bm\chi_n$ of MU $n \in \mathcal{N}_i$.

%-----------------------------------------------------------
We can now specify the number of needed channels by a SP $i \in \mathcal{I}$ in the area of a BS $b \in \mathcal{B}$ as $C_{b, i} = \sum_{\{n \in \mathcal{N}_i: L_n \in \mathcal{L}_b\}} z_n $ and the valuation of obtaining $\mathbf{C}_i = (C_{b, i}: b \in \mathcal{B})$ across the whole service area as
\begin{align}\label{valu}
      \nu_i
  & = \frac{1}{1 - \gamma} \cdot \sum\limits_{n \in \mathcal{N}_i} \alpha_n \cdot \mathds{U}_n(\bm\chi_n)                   \nonumber\\
  & - \frac{\gamma}{1 - \gamma} \cdot \sum_{s_i' \in \mathcal{S}_i}
      \mathbb{P}\!\left(s_i' | s_i, \mathds{1}_{\left\{\sum_{b \in \mathcal{B}} C_{b, i} > 0\right\}}\right) \cdot
      \mathds{U}_i\!\left(s_i'\right),
\end{align}
which together constitute a bid $\tilde{\pi}_{i, (\mathrm{c})}^*(\tilde{\bm\chi}_i) = \bm\beta_i \triangleq (\nu_i, \mathbf{C}_i)$ of SP $i$ in $\tilde{\bm\chi}_i \in \tilde{\mathcal{X}}_i$, where $z_n$ is given by (\ref{whetChan})
\begin{figure*}[!t]
\vspace{-.8cm}
\begin{align}\label{whetChan}
 &   z_n =                                                                                                                                  \nonumber\\
 &   \underset{z \in \{0, 1\}}{\arg\max} \left\{(1 - \gamma) \cdot
     U_n\!\left(\bm\chi_n, z, \pi_{n, (\mathrm{t})}^*(\bm\chi_n), \pi_{n, (\mathrm{p})}^*(\bm\chi_n)\right)
   + \gamma \cdot \sum_{\bm\chi_n' \in \mathcal{X}}
     \mathbb{P}\!\left(\!\bm\chi_n' | \bm\chi_n, z, \pi_{n, (\mathrm{t})}^*(\bm\chi_n), \pi_{n, (\mathrm{p})}^*(\bm\chi_n)\right) \cdot
     \mathds{U}_n(\bm\chi_n')\right\}
\end{align}
\vspace{-.2cm}
\hrule
\end{figure*}
and $\mathds{1}_{\{\Xi\}}$ equals $1$ if the condition $\Xi$ is satisfied and $0$, otherwise.

\subsection{Learning Optimal Abstract Control Policy}
\label{loacp}

We can easily find that at a current scheduling slot, $\bm\beta_i$ of a SP $i \in \mathcal{I}$ needs $(s_i, \mathbb{P}(s' | s, \iota - 1))$ and $(\mathds{U}_n(\bm\chi_n), z_n, L_n)$ from each subscribed MU $n \in \mathcal{N}_i$, where $s' \in \mathcal{S}_i$ and $\iota \in \{1, 2\}$.
We propose that SP $i$ maintains over the slots a three-dimensional table $\mathbf{Y}_i^k$ of size $S_i \cdot S_i \cdot 2$.
Each entry $y_{s, s', \iota}^k$ in $\mathbf{Y}_i^k$ represents the number of transitions from $s_i^{k - 1} = s$ to $s_i^k = s'$ when $\phi_i^{k - 1} = \iota - 1$ up to slot $k$.
$\mathbf{Y}_i^k$ is updated using the channel auction outcomes.
Then, we estimate the abstract network state transition probability at a slot $k$ as
\begin{align}
    \mathbb{P}\!\left(s_i^k = s' | s_i^{k - 1} = s, \phi_i^{k - 1} = \iota - 1\right) =
    \frac{y_{s, s', \iota}^k}{\sum\limits_{s'' \in \mathcal{S}_i} y_{s'', s', \iota}^k},
\end{align}
based on which $\mathds{U}_i(s_i)$, $\forall s_i \in \mathcal{S}_i$ is learned via (\ref{paymUpda})
\begin{figure*}[!t]
\vspace{-.6cm}
\begin{align}\label{paymUpda}
   \mathds{U}_i^{k + 1}(s_i) =
   \left\{\!\!
   \begin{array}{l@{~}l}
     \left(1 - \zeta^k\right) \cdot \mathds{U}_i^k(s_i) + \zeta^k \cdot \left((1 - \gamma) \cdot \tau_i^k +
            \gamma \cdot \displaystyle\sum_{s_i^{k + 1} \in \mathcal{S}_i} \mathbb{P}\!\left(s_i^{k + 1} | s_i, \phi_i^k\right) \cdot
            \mathds{U}_i^k\!\left(s_i^{k + 1}\right)\right),                                                    & \mbox{if } s_i = s_i^k   \\
     \mathds{U}_i^k(s_i),                                                                                       & \mbox{otherwise}
   \end{array}
   \right.
\end{align}
\vspace{-.2cm}
\hrule
\end{figure*}
with $\zeta^k \in [0, 1)$ being the learning rate.
(\ref{paymUpda}) converges if $\sum_{k = 1}^\infty \zeta^k = \infty$ and $\sum_{k = 1}^\infty (\zeta^k)^2 < \infty$ \cite{Rich98}.

Without a priori statistics of MU mobility and computation task as well as packet arrivals, $Q$-learning \cite{Rich98} finds $\mathds{U}_n(\bm\chi_n)$ for each MU $n \in \mathcal{N}$ by defining the right-hand-side of (\ref{MTBell}) as the optimal state action-value function $Q_n: \mathcal{X} \times \{0, 1\} \times \mathcal{A} \times \mathcal{W} \rightarrow \mathds{R}$.
In turn, we arrive at
\begin{align}\label{stat_acti_q2}
  \mathds{U}_n(\bm\chi_n) =
  \max_{\varphi_n, R_{n, (\mathrm{t})}, R_{n, (\mathrm{p})}} Q_n\!\left(\bm\chi_n, \varphi_n, R_{n, (\mathrm{t})}, R_{n, (\mathrm{p})}\right),
\end{align}
where an action $(\varphi_n, R_{n, (\mathrm{t})}, R_{n, (\mathrm{p})})$ under a current local state $\bm\chi_n$ consists of the channel allocation, computation offloading and packet scheduling decisions.
The tabular nature in representing $Q$-function values makes the conventional $Q$-learning not readily applicable.
In our considered network, the sizes of $\mathcal{X}$ and action space $\{0, 1\} \times$ $\mathcal{A} \times \mathcal{W}$ are calculated as $|\mathcal{L}|^2 \cdot (1 + A_{(\mathrm{t})}^{(\max)}) \cdot (1 + W^{(\max)})$ and $2 \cdot (1 + A_{(\mathrm{t})}^{(\max)}) \cdot (1 + W^{(\max)})$, resulting in an extremely slow learning process.

The success of a deep neural network in modelling the $Q$-function inspires us to adopt a deep reinforcement learning (DRL) method \cite{Mnih15}.
We can then approximate the $Q$-function by a double deep $Q$-network (DQN) \cite{Hass16}.
Mathematically, $Q_n(\bm\chi_n, \varphi_n, R_{n, (\mathrm{t})}, R_{n, (\mathrm{p})}) \approx Q_n(\bm\chi_n, \varphi_n, R_{n, (\mathrm{t})}, R_{n, (\mathrm{p})}; \bm\theta_n)$, $\forall n \in \mathcal{N}$, where we encapsulate in $\bm\theta_n$ the set of parameters that are associated with the DQN of a MU $n$.
During the DRL process, each MU $n \in \mathcal{N}_i$ of a SP $i \in \mathcal{I}$ is assumed to be equipped with a finite replay memory to store the latest $M$ historical experiences, namely, $\mathcal{M}_n^k = \{\mathbf{m}_n^{k - M + 1}, \cdots, \mathbf{m}_n^k\}$, where each experience $\mathbf{m}_n^{k'} = (\bm\chi_n^{k'}, (\varphi_n^{k'}, R_{n, (\mathrm{t})}^{k'}, R_{n, (\mathrm{p})}^{k'}),$ $U_n(\bm\chi_n^{k'}, \varphi_n^{k'}, R_{n, (\mathrm{t})}^{k'}, R_{n, (\mathrm{p})}^{k'}), \bm\chi_n^{k' + 1})$ happens at the transition between two consecutive scheduling slots $k'$ and $k' + 1$.
To perform experience replay \cite{Lin92}, MU $n$ randomly samples a mini-batch $\mathcal{O}_n^k \subseteq \mathcal{M}_n^k$ to train the DQN parameters using the loss function in (\ref{lossFunc}),
\begin{figure*}[!t]
\vspace{-.6cm}
\begin{align}\label{lossFunc}
 &   \textsf{LOSS}_n\!\left(\bm\theta_n^k\right)
   = \textsf{E}_{\left(\bm\chi_n, \left(\varphi_n, R_{n, (\mathrm{t})}, R_{n, (\mathrm{p})}\right),
     U_n\left(\bm\chi_n, \varphi_n, R_{n, (\mathrm{t})}, R_{n, (\mathrm{p})}\right), \bm\chi_n'\right) \in \mathcal{O}_n^k}\Bigg[\Bigg(
     (1 - \gamma) \cdot U_n(\bm\chi_n, \varphi_n, R_{n, (\mathrm{t})}, R_{n, (\mathrm{p})})~+                                       \nonumber\\
 &   \gamma \cdot Q_n\!\left(\bm\chi_n', \underset{\varphi_n', R_{n, (\mathrm{t})}',
     R_{n, (\mathrm{p})}'}{\arg\max} Q_n\!\left(\bm\chi_n', \varphi_n', R_{n, (\mathrm{t})}',
     R_{n, (\mathrm{p})}'; \bm\theta_n^k\right); \bm\theta_{n, -}^k\right) -
     Q_n\!\left(\bm\chi_n, \varphi_n, R_{n, (\mathrm{t})}, R_{n, (\mathrm{p})}; \bm\theta_n^k\right)\Bigg)^2\Bigg]
\end{align}
\vspace{-.2cm}
\hrule
\vspace{-.6cm}
\end{figure*}
where $\bm\theta_n^k$ and $\bm\theta_{n, -}^k$ are, respectively, the DQN parameters at a scheduling slot $k$ and a certain previous scheduling slot before slot $k$.

\section{Numerical Experiments}
\label{simu}

This section conducts numerical experiments based on TensorFlow \cite{Abad16} to quantify the performance of the derived DRL-based scheme for multi-tenant cross-slice resource orchestration with secrecy preserving in a software-defined RAN.
We set up an experimental network with $4$ BSs being placed at equal distance $1$ Km apart in the centre of a $2\times2$ Km$^2$ square service area \cite{Chen15}.
The entire area is divided into $1600$ locations with each of 50$\times$50 m$^2$.
The average channel gains for a MU $n \in \mathcal{N}$ at the location $L_{n, (\mathrm{u})}^k \in \mathcal{L}_b$ covered by a BS $b \in \mathcal{B}$ during a slot $k$ are given by $h_{(\mathrm{u})}(L_{n, (\mathrm{u})}^k) = H_0 \cdot (\xi_0/\xi_{b, n}^k)^4$ and $h_{(\mathrm{e})}(L_{n, (\mathrm{u})}^k, L_{(\mathrm{e})}^k) = H_0 \cdot (\xi_0/\xi_{n, (\mathrm{e})}^k)^4$, where $H_0 = -40$ dB is the path-loss constant, $\xi_0 = 2$ m is the reference distance, while $\xi_{b, n}^k$ and $\xi_{n, (\mathrm{e})}^k$ are the distances between MU $n$ and BS $b$ as well as the eavesdropper \cite{Mao16}.
The mobilities of all MUs as well as the eavesdropper and the computation task arrivals of all MUs are independently and randomly generated.
The packet arrivals follow a Poisson arrival process with average rate $\lambda$ (in packets/slot).
For the utility function in (\ref{util}), we select $U_n^{(1)}(W_n^{k + 1}) = \exp\{- W_n^{k + 1}\}$, $U_n^{(2)}(D_n^k) = \exp\{- D_n^k\}$, $U_n^{(3)}(P_{n, (\mathrm{CPU})}^k) = \exp\{- P_{n, (\mathrm{CPU})}^k\}$ and $U_n^{(4)}(P_{n, (\mathrm{tr})}^k) = \exp\{- P_{n, (\mathrm{tr})}^k\}$.
We design for each MU a DQN with $2$ hidden layers with each consisting of $16$ neurons.
Other parameter values used in the experiments are listed in Table \ref{tabl2}.

\begin{table}[t]
  \caption{Parameter values in experiments.}\label{tabl2}
        \begin{center}
        \begin{tabular}{c|c}
              \hline
              Parameter                                                     & Value                                             \\\hline
              Set of SPs                      $\mathcal{I}$                 & $\{1, 2, 3\}$                                     \\\hline
              Set of BSs                      $\mathcal{B}$                 & $\{1, 2, 3, 4\}$                                  \\\hline
              Number of MUs                   $|\mathcal{N}_i|$             & $6$, $\forall i \in\mathcal{I}$                   \\\hline
              Channel bandwidth               $\eta$                        & $500$ KHz                                         \\\hline
              Noise power spectral density    $\sigma^2$                    & $-174$ dBm/Hz                                     \\\hline
              Scheduling slot duration        $\delta$                      & $10^{-2}$ second                                  \\\hline
              Discount factor                 $\gamma$                      & $0.9$                                             \\\hline
              Utility price                   $\alpha_n$                    & $1$, $\forall n\in\mathcal{N}$                    \\\hline
              Packet size                     $\mu_{(\mathrm{p})}$          & $3000$ bits                                       \\\hline
              Maximum transmit power          $\Omega^{(\max)}$             & $3$ Watts                                         \\\hline
              Weight of energy consumption    $\ell_n$                      & $3$, $\forall n\in\mathcal{N}$                    \\\hline
              Maximum queue length            $W^{(\max)}$                  & $10$ packets                                      \\\hline
              Maximum task arrivals           $A_{(\mathrm{t})}^{(\max)}$   & $5$ tasks                                         \\\hline
              Input data size                 $\mu_{(\mathrm{t})}$          & $5000$ bits                                       \\\hline
              CPU cycles per bit              $\vartheta$                   & $737.5$                                           \\\hline
              CPU-cycle frequency             $\varrho$                     & $2$ GHz                                           \\\hline
              Effective switched capacitance  $\varsigma$                   & $2.5 \cdot 10^{-28}$                              \\\hline
              Exploration probability         $\epsilon$                    & $0.001$                                           \\\hline
              Replay memory size              $M$                           & $5000$                                            \\\hline
              Mini-batch size                 $|\mathcal{O}_n^k|$           & $200$, $\forall n \in \mathcal{N}$, $\forall k$   \\\hline
              Activation function                                           & Tanh \cite{Jarr09}                                \\\hline
              Optimizer                                                     & Adam \cite{King15}                                \\
              \hline
        \end{tabular}
    \end{center}
\end{table}

For comparison purpose, three baseline schemes are developed and simulated, namely,
\begin{enumerate}
  \item Channel-aware control policy (Baseline 1) -- At the beginning of each slot $k$, the need of getting one channel at a MU $n \in \mathcal{N}$ is evaluated by $H_{n, (\mathrm{u})}^k - H_{(\mathrm{e})}^k$;
  \item Queue-aware control policy (Baseline 2) -- Each MU calculates the preference between having one channel or not using a predefined threshold of the queue length;
  \item Random control policy (Baseline 3) -- This policy randomly generates the value of obtaining one channel for each MU at the beginning of each slot.
\end{enumerate}
With the three baselines, after the centralized channel allocation by the SDN-orchestrator at the beginning of each slot, a MU proceeds to offload a random number of computation tasks and schedule a maximum feasible number of data packets if being assigned a channel.

\begin{figure}[t]
    \centering
    \includegraphics[width=19pc]{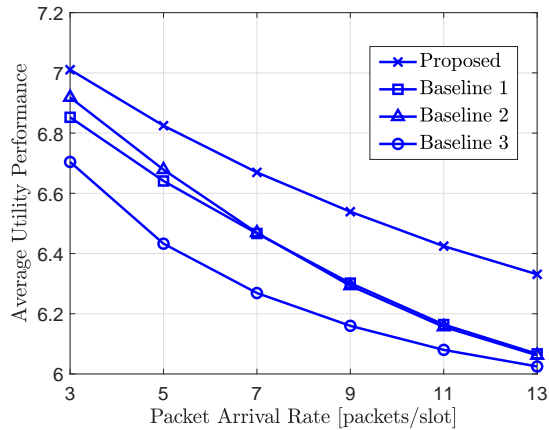}
    \caption{Average utility performance per MU across the learning procedure versus average packet arrival rates.}
    \label{simu01}
\end{figure}

\begin{figure}[t]
    \centering
    \includegraphics[width=19pc]{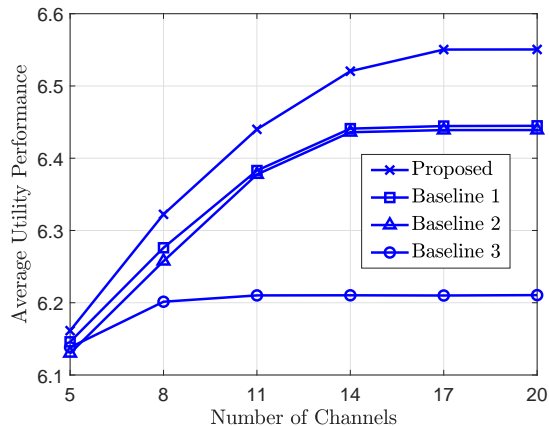}
    \caption{Average utility performance per MU across the learning procedure versus numbers of channels.}
    \label{simu02}
\end{figure}

We first demonstrate the average utility performance per MU per scheduling slot achieved from the proposed DRL-based scheme and the three baselines under different average packet arrival rates.
In this experiment, we assume that $J = 11$ channels are shared among the MUs for the access to the computation and communication slices.
The results are depicted in Fig. \ref{simu01}, from which we can observe that the proposed scheme achieves a significant performance gain.
However, the average utility performance deceases as the average number of random data packet arrivals increases.
The reason behind is that in order to ensure secrecy, more data packet arrivals lead to larger queue length, more packet drops and higher energy consumption across the MUs.
Then in Fig. \ref{simu02}, we exhibit the average utility performance versus the number of channels, where the average packet arrival rate is fixed to be $\lambda = 8$.
More channels available in the system provide more opportunities for the MUs to transmit the data of computation tasks to be offloaded and scheduled packets.
Hence better average utility performance can be expected by the MUs.
When there are sufficient channels in the network, the data transmissions of all MUs with secrecy preserving can be fully satisfied.
Both experiments show that the proposed scheme outperforms the three baselines.

\section{Conclusions}
\label{conc}

In this paper, we investigate the problem of non-cooperative multi-tenant cross-slice resource orchestration with secrecy preserving in a software-defined RAN, which is formulated as a $\mathcal{SG}$.
To alleviate private information exchange among the competing SPs, we approximate the $\mathcal{SG}$ by a $\mathcal{AG}$.
Each SP is thus able to behave independently only with the local information.
We observe that the decisions of the channel auction and the computation offloading as well as packet scheduling are sequentially made.
This motivates us to linearly decompose the per-SP single-agent MDP, which greatly simplifies the decision making process at a SP.
We propose a DRL-based scheme to find the optimal abstract control policies.
Numerical experiments showcase that the performance achieved from our scheme outperforms the other baselines.

\end{document}